\documentclass{aastex}
\usepackage{spr-astr-addons}
\usepackage{url}

\RequirePackage{color}

\usepackage{graphicx}

\def \beq {\begin{equation}}
\def \eeq {\end{equation}}

\begin{document}

\title{Searching for curvature pion radiation from protons in strongly magnetized pulsars}
\slugcomment{Not to appear in Nonlearned J., 45.}
\shorttitle{Curvature pion radiation from protons}
\shortauthors{Fregolente and Saa}

\author{Douglas Fregolente\altaffilmark{1}}
\affil{Universidade Federal de Campina Grande, Centro de Forma\c c\~ao de Professores, Cajazeiras, PB,  Brazil}
\and
\author{Alberto Saa\altaffilmark{2}}
\affil{Departamento de Matem\'atica Aplicada,
UNICAMP,
  Campinas, SP, Brazil}

\altaffiltext{1}{douglasfregolente@cfp.ufcg.edu.br}
\altaffiltext{2}{asaa@ime.unicamp.br}

\begin{abstract}
By using rather conservative estimates based on the simplest polar cap model,  we search the ATNF Pulsar Catalogue  for strongly magnetized stars that could accelerate relativistic protons up to the curvature pion production threshold.
The best candidate turns out to be the 16 ms pulsar J0537-6910, but the  corresponding characteristic parameter $\chi=a/m_p$ is yet  too
small     to give origin to  observable signals.
 We show that, for pulsars with period $P\approx 1\,$ms, a surface polar magnetic field $B \approx 10^{12}$G is required in order to induce detectable curvature pion radiation from accelerated protons in the magnetosphere. Some other emission processes are also considered.
\end{abstract}


\section{Introduction}
Processes related to the
  emission of pions by accelerated protons like, for instance, $p^+\stackrel{a}{\to} p^+\pi^0$,
  have been considered in the literature since the sixties\citep{GZh,Zh,GS,Ritus}.
  The present available observational data on
  strongly magnetized
 astrophysical objects could provide an exciting scenario to test theoretical
 predictions involving such kind of inertially forbidden processes. The relevant observer-independent  parameter characterizing  these phenomena
  is the dimensionless quantity $\chi = a/m_p$, where $a$ and $m_p$ stand, respectively, for the proton proper acceleration and rest mass. (Unless otherwise stated, natural unities are adopted through  this work.) For $\chi \ll 1$, one can employ a semiclassical approach where the proton is described by a classical current while the pion is considered as a fully quantized field. This corresponds to the so called no-recoil approximation, for which  many analytical formulas are available. (For some recent works on the subject, see, for instance, \citep{Herpay:2007vu,Herpay:2008gd} and \citep{Fregolente:2006xj,Fregolente:2007qw} for the electromagnetic and gravitational cases, respectively.)
  In particular, we expect a strong suppression of the pion emission   for $\chi\to 0$   since such a process is well known to be inertially forbidden.
  For $\chi \gg 1$, a full quantized treatment is mandatory\citep{Ritus,Berezinsky:1995bf}. However,   numerical evidences suggest that some  appropriate limits of certain results obtained with the no-recoil approximation can still be considered  as good estimates\citep{Tokuhisa:1999wg}. For the case of the usual
   photon  synchrotron radiation, for instance, quantum
effects were carefully considered\citep{Erber:1966vv} and the semiclassical
approach was found to be accurate  within a few percent in the limit $\chi \gg 1$. Notice that, in the cases where the proton acceleration is caused by strong magnetic fields, the parameter $\chi$ can be written in a similar way to the usual  synchrotron radiation:  $\chi=\gamma B/B_{\rm cr}$, where $\gamma$ is the   Lorentz factor, $\gamma B$ is the magnetic field in the instantaneous reference frame of the proton, and  $B_{\rm cr} = m_p^2/e \approx 1.5 \times 10^{20}$ G is a critical magnetic field strength, denoting the limit of the validity of the no-recoil approximation.

Thus, we have basically two regimes where the details of the pion emission by accelerated protons are known:  the no-recoil regime $a \ll m_\pi$, where $m_\pi \approx 135$ MeV stands for the $\pi^0$ mass, and the ultra-accelerated one, $a \gg m_p \approx 938$ MeV. The intermediate acceleration range corresponding to $m_\pi < a < m_p$, which could encompass interesting astrophysical applications, is still open for debate. It is clear, however, that for such acceleration values,   the pion emission channel will certainly be excited.
In fact, the $\pi^0$ radiation process starts to become competitive with
the usual photon  synchrotron radiation only for   $\chi \approx m_\pi/m_p$\citep{Tokuhisa:1999wg,Herpay:2008gd}.

The dominant decay mode for $\pi^0$ is the electromagnetic channel, giving origin to two photons\citep{PDG}. From the observational point of view,   charged pions are more promising since they decay preferentially by the weak  channel in muons and neutrinos\citep{PDG}. Processes like  $p^+\stackrel{a}{\to} n \pi^+$
have also been considered\citep{Herpay:2007vu,Fregolente:2006xj}, and they
have essentially the same two regimes according to the value of the parameter
$\chi$. We notice, however, that the semiclassical analysis of such processes in magnetic fields is rather tricky  since the neutron will not be accelerated  and, consequently, will  follow a different trajectory from the proton, implying that the initial ($p^+$) and final ($n$)  currents are qualitatively different, in sharp contrast to the gravitational case\citep{{Fregolente:2006xj,Fregolente:2007qw}}. On the other hand,  cases involving vector mesons  like
$p^+\stackrel{a}{\to} p^+\rho $, with $m_\rho = 770$ MeV, as those ones considered in \cite{Tokuhisa:1999wg}, are analogous to the neutral $\pi^0$ case.

In the present work, we search the ATNF Pulsar Catalogue\citep{Manchester:2004bp} for strongly magnetized pulsars that could give rise to appreciable values for the parameter $\chi$. We employ rather conservative estimates based on the simplest polar cap model for pulsars. The highest value of $\chi$ we found corresponds to the 16 ms pulsar J0537-6910. Unfortunately, the obtained value   $\chi \approx 10^{-6}$ is too small in order to give origin to observable signals. Nevertheless,
our estimates reveal that
 a pulsar with period $P=1\,$ms and surface polar magnetic field $B =  10^{12}$G would suffice to induce
  detectable pion radiation from accelerated protons. Our results can be applied also to the weak interaction channel emission, as in the process $p^+\stackrel{a}{\to} ne^+\nu$. The threshold in this case correspond to the electron mass $m_e \approx 0.5$ MeV.
 The $\chi$ values   for the known pulsars, however, are  not enough to excite  this emission channels either. For a $P=1\,$ms pulsar, a magnetic field $B \approx 4\times 10^{10}$G would be required to induce the  weak interaction emissions.

\section{Particle acceleration in the polar cap model}

We use here the simplest polar cap model for pulsars. (For a comprehensive  review on the subject, see \cite{Harding:2006qn}.)
 We recall that, in the magnetosphere of   strongly magnetized pulsars, one expects   an electric field component parallel to the magnetic field, which could
accelerate charged particles up to relativistic velocities along
the magnetic field lines near the polar region of the star. Pulsars with polar caps for which
$\mathbf{\Omega}\cdot\mathbf{B}<0$, where $\mathbf{\Omega}$ and $\mathbf{B}$ stand, respectively, for the rotation and the magnetic field vectors, could effectively accelerate protons away from the polar region surface.
In the so called inner
acceleration region\citep{Harding:2006qn}, the accelerated particles are expected to radiate away
their transverse energy so efficiently that they remain effectively constrained to move along the curved magnetic field lines  with constant energy, corresponding to some
Landau level.

The curvature radius $R_c$  of the polar magnetic field lines can be
approximated by\citep{Harding:1998ma}
\beq
\label{Rc}
R_c^2 \approx \frac{8G^2}{9\pi} rP,
\eeq
where  $r$ is the radius   of the star, $P$ is the rotation period,  and $G$ is a curved space correction term, which tends to increase
$R_c$ by 25\% - 30\% when compared to the flat space formula. We assume the
flat space value ($G=1$)
since we are going to estimate the maximum possible centripetal acceleration
attained by protons.
The curvature radius (\ref{Rc}) does indeed correspond to the
last open magnetic field line of the pulsar.
For a relativistic particle following a circular trajectory with radius $R_c$,
the proper centripetal acceleration is given by $a\approx \gamma^2/R_c$. Assuming that the  kinetic energy $\epsilon_p$ of the accelerated proton moving along the last open magnetic field line was obtained from the parallel electric field, one has
\beq
\label{gamma}
\epsilon_p = \gamma m_p =  \eta e\Delta\phi,
\eeq
where $\eta$ accounts to the efficiency of the  acceleration mechanism\citep{Zhang:1999ua,meszaros} and
\beq
\Delta\phi = 2\pi^2\frac{Br^3}{P^2}
\eeq
is the
Goldreich-Julian
maximum potential drop near the star surface\citep{Goldreich:1969sb}, with $B$ being the pulsar surface  polar magnetic field. The maximum efficiency is
expect to be as high as $0.85$ \citep{Zhang:1999ua}, but here
we suppose total efficiency, $\eta =1$. Finally,
the maximum centripetal acceleration of such protons leaving the polar cap
 along the last open magnetic field line can be approximated by
\beq
\label{acce}
\frac{a}{m_p} \approx 3\sqrt{2\pi^9}m_p\left(\frac{ B}{B_{\rm cr}} \right)^2
\sqrt{\frac{r^{11}}{P^9}}.
\eeq
For our purposes here,
it is conveniente to introduce the usual dimensionless quantities $P_{-3}$  and $B_{12}$ defined, respectively, by
 $P=P_{-3}\times 10^{-3}$s  and $B = B_{12}\times 10^{12}$G.
Assuming a typical radius $r =10^4\,$m for the pulsar, one has finally
\beq
\label{chi}
\chi  = \frac{a}{m_p} \approx 0.35    B_{12}^2 P_{-3}^{-\frac{9}{2}}.
\eeq
The total energy acquired by the proton is another important quantity in our analysis. From (\ref{gamma}), we have
\beq
\gamma = 2\pi^2 m_p \frac{B}{B_{\rm cr}} \frac{r^3}{P^2},
\eeq
leading that
\beq
\label{gamma1}
\gamma = 7\times 10^{9}B_{12} P^{-2}_{-3}
\eeq
for stars of the typical size.

\section{Results}

We evaluate $\chi$ as given by (\ref{chi}) for all the 1700 pulsars in ATNF catalogue\citep{Manchester:2004bp} with known $P$ and $B$ values. Table \ref{table1} shows the ten pulsars with the highest values of  $\chi$.
\begin{table}
  \begin{tabular}{  l   r   r   r   r   }
    \\\hline
    $\quad$ Pulsar $\quad $& $\quad $$\quad$$P_{-3}$  & $\quad$$B_{12}$  & $\quad$$\chi_{-7}$ & $\quad$$\quad$$\gamma_7\,$     \\ \hline
    J0537-6910 & 16.12 & 0.93  & 11.0  & 2.49 \\
    B0531+21   & 33.08 & 3.78  & 7.26  & 2.42 \\
    B0540-69   & 50.50 & 4.98  & 1.88  & 1.37 \\
    J1813-1749 & 44.70 & 2.65  & 0.92  & 0.93 \\
    J1400-6325 & 31.18 & 1.11  & 0.82  & 0.80 \\
    J1747-2809 & 52.15 & 2.88  & 0.54  & 0.74 \\
    J1833-1034 & 61.87 & 3.58  & 0.39  & 0.65 \\
    J0205+6449 & 65.59 & 3.61  & 0.30  & 0.59 \\
    J2229+6114 & 51.62 & 2.03  & 0.28  & 0.53 \\
    J1617-5055 & 69.36 & 3.10  & 0.17  & 0.45 \\
    \hline
  \end{tabular}
  \caption{\label{table1} Top ten  $\chi$ parameters,
  calculated according
  to (\ref{chi}), for the 1700 objects of the ATNF Pulsar Catalogue\citep{Manchester:2004bp} with known $P$ and $B$ values. The last column is the total proton energy given by (\ref{gamma1}).
  The pulsar period and magnetic field are, respectively,
 $P=P_{-3}\times 10^{-3}$s  and $B = B_{12}\times 10^{12}$G. Analogously,
 we have $\chi=\chi_{-7}\times 10^{-7}$ and $\gamma = \gamma_7\times 10^7$.}
\end{table}
As one can see, these values are too small to give rise to detectable signals, since the pion emission channel is effectively excited only for accelerations such that $\chi \approx m_\pi/m_p\approx 0.14$. Nevertheless, the proton can indeed reach considerable high energies, as the last column of Table  \ref{table1} shows. For the J0537-6910 pulsar, for instance, the proton can reach an energy $\epsilon_p\approx 2.3\times 10^5\,$GeV.
We notice that for the pulsar with the highest surface magnetic field (J1808-2024, $B\approx 2\times 10^{15}\,$G), one has only $\chi\approx 5.2\times 10^{-12}$, since its period is considerable large ($P\approx 7.6\,$s).

\begin{figure*}[t]
\begin{center}
  \includegraphics[width=0.75\linewidth]{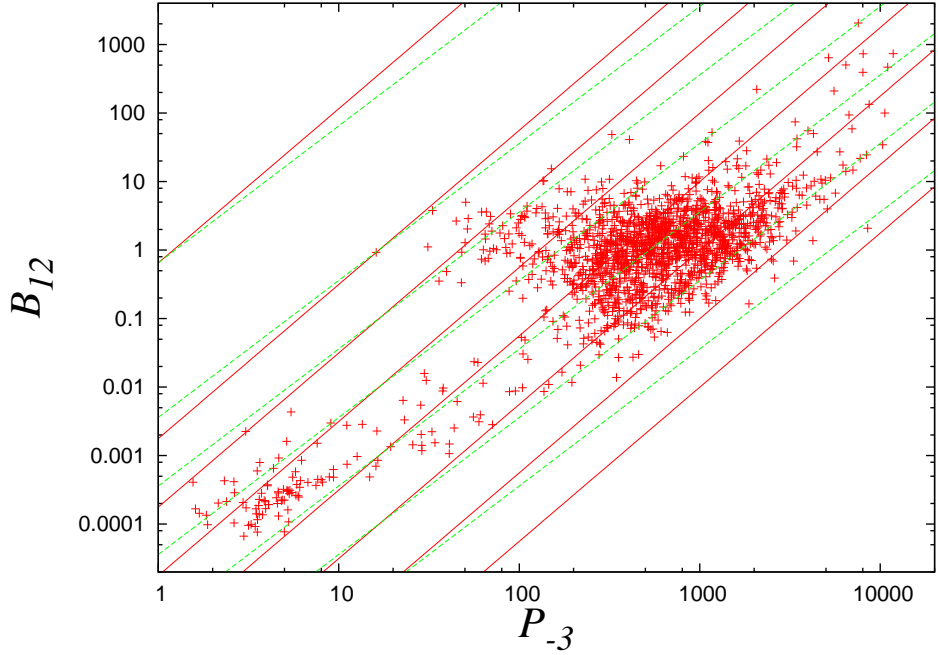}
\end{center}
\caption{The values $(P_{-3},B_{12})$ for the 1700 objects in the ATNF Pulsar Catalogue\citep{Manchester:2004bp} with known period $P = P_{-3}\times 10^{-3}\,$s and surface magnetic field $B=B_{12}\times 10^{12}\,$G. The uppermost red (solid) and green (dashed) lines correspond, respectively, to the curves of constant $\chi\approx m_\pi/m_p$ (\ref{chi}) and $\gamma \approx 4.5\times 10^{9}$ (\ref{gamma1}). The other red (solid) and green (dashed) lines correspond, respectively, to the lines with $\chi\approx 10^{-7}, 10^{-9},\dots,10^{-14}$ and
with $\gamma \approx 10^7,10^6,\dots,10^2.$ All the pulsars
in the ATNF catalogue   are in a region of the plane $(P_{-3},B_{12})$ from where no detectable signals are expected from curvature pion radiation emitted by accelerated protons in the magnetosphere.
  }
\label{fig1}
\end{figure*}

Figure \ref{fig1} show the     $(P_{-3},B_{12})$ pairs  for the considered pulsars.
All the known pulsars are contained in the regions bounded by
the curves (\ref{chi}) and (\ref{gamma1}) with, respectively, $\chi=10^{-6}$ and
$\gamma =2\times 10^7$ (upper) and $\chi=10^{-14}$ and $\gamma = 2\times 10^1$ (bottom).
The small values obtained for $\chi$ are, of course, a consequence of the large proton mass $m_p$. Had we considered electrons instead of protons, we would have $B_{\rm cr}\approx 4.3\times 10^{13}$G, and the parameter $\chi$ would be 10 orders of magnitude as larger. The regime $\chi\approx 1$ can   be easily  reached for electrons, explaining the rich physics  of electrons in typical pulsar magnetospheres.
Processes involving muons in circular
orbits, like $\mu^-\to e^-\bar{\nu}_e\nu_\mu$ for instance,  have also been investigated recently in the gravitational context \citep{Fregolente:2007qw}. The muon mass is $m_\mu = 107$ MeV, implying   a muon acceleration parameter $\chi = a/m_\mu\approx 7.4\times 10^{-4}$ for the pulsar J0537-6910. This is not enough to lead to any appreciable deviation of the muon half life as considered in \cite{Fregolente:2007qw}.

Since we have $\chi \ll 1$, we can apply safely the formulas obtained in the no-recoil approximation for the process $p^+\stackrel{a}{\to} p^+\pi^0$. The total energy emitted by the pion synchrotron radiation, or in other words, the
proton energy loss rate due to curvature pion radiation  is given by\citep{Tokuhisa:1999wg}
\beq
{\cal R} = \frac{g^2}{\sqrt{3}} m_\pi m_p \chi \exp\left({-\frac{\sqrt{3}}{\chi}\frac{m_\pi}{m_p}}\right)
\eeq
in the limit $\chi \ll 1$, where $g^2\approx 14$ is the strong coupling constant. The so called cooling time
$t_\pi = \epsilon_p/{\cal R}$
is an important parameter   corresponding to the time scale for which the process becomes relevant. For our case, we have
\beq
\label{tpi}
t_\pi = 6 \times 10^{-25}  \gamma \chi^{-1} \exp\left( 0.25\chi^{-1} \right)\,{\rm s},
\eeq
for stars of the typical size, with $\chi$ and $\gamma$ given, respectively, by (\ref{chi}) and (\ref{gamma1}).
 The exponential term in (\ref{tpi})   gives rise to
unreasonable high  values of $t_\pi$ for the pulsars in the ATNP catalogue when compared, for instance, with the acceleration time $t_a$ \citep{Herpay:2008gd}.
From Figure \ref{fig1},
we can also infer
 the location of the  curves with constant $t_\pi$  in the  $(P_{-3},B_{12})$ plane.
Notice that the curves of constant $\chi$ and $\gamma$ are close in the log-log plot, since their inclination are, respectively $9/4$ and 2. Regions of approximately constant $\chi$ and $\gamma$ would give rise, according to (\ref{tpi}), to approximately constant $t_\pi$. So, the curve of constant $t_\pi$ passing through the point $(P_{-3},B_{12})$  will
be quite close to the lines of constant $\chi$ and $\gamma$ intercepting at the same point.  For the
the values of $\chi$ and $\gamma$ corresponding to the upper bound of the known pulsars,
we would have unreasonable highs values for $t_\pi$. This is, of course, a consequence of the strong suppression of the pion radiation emitted by quasi-inertial protons.
It is clear   that no currently known pulsar could origin detectable signals
of the curvature pion radiation emitted  from accelerated protons in the
magnetosphere.

\acknowledgments
This work was supported by FAPESP and CNPq.

\nocite{*}
\bibliographystyle{spr-mp-nameyear-cnd}

\begin{thebibliography}{}

\bibitem[Ginzburg \& Zharkov(1965)]{GZh} Ginzburg, V. L., \& Zharkov, G. F. 1965, Sov. Phys. JETP, 20,
1525.
\bibitem[Zharkov(1965)]{Zh} Zharkov, G. F. 1965, Sov. J. Nucl. Phys., 1, 120.

\bibitem[Ginzburg \& Sirovatski(1965)]{GS} Ginzburg, V. L., \& Sirovatski,  S. I., Ann. Rev. Astron. Astrophys., 3, 297.

\bibitem[Ritus(1985)]{Ritus} Ritus, V. I. 1985, J. Sov. Laser Research, 6, 497.

\bibitem[Herpay \& Patkos(2008)]{Herpay:2007vu}
  Herpay, T., \& Patkos, A. 2008,
  J.\ Phys.\ G, 35, 025201.

\bibitem[Herpay et al. (2008)]{Herpay:2008gd}
  Herpay, T., Razzaque, S., Patkos, A., \& Meszaros, P. 2008,
  JCAP, 0808, 025.


\bibitem[Fregolente, Matsas \& Vanzella(2006)]{Fregolente:2006xj}
  Fregolente, D., Matsas, G. E. A., \& Vanzella, D. A. T. 2006,
  Phys.\ Rev.\  D, 74, 045032.

\bibitem[Fregolente \& Saa(2008)]{Fregolente:2007qw}
  Fregolente, D., \& Saa, A. 2008,
  Phys.\ Rev.\  D, 77, 103010.
 

\bibitem[Berezinsky, Dolgov \& Kachelriess(1995)]{Berezinsky:1995bf}
 Berezinsky, V., Dolgov, A., \& Kachelriess, M. 1995,
  Phys.\ Lett.\  B, 351, 261.

\bibitem[Tokuhisa \& Kajino(1999)]{Tokuhisa:1999wg}
  Tokuhisa, A., \& Kajino, T. 1999,
  Ap.J.Lett., 525, 117.

\bibitem[Erber(1966)]{Erber:1966vv}
  Erber, T. 1966,
  Rev.\ Mod.\ Phys.,  38, 626.

\bibitem[Nakamura et al.(2010)]{PDG} Nakamura, K. {\em et al.} 2010 (Particle Data Group),
J. Phys. G, 37, 075021 [http://pdg.lbl.gov]

\bibitem[Manchester et al.(2004)]{Manchester:2004bp}
  Manchester, R. N., Hobbs, G.~B., Teoh, A., \& Hobbs, M. 2005,
  Astron.\ J., 129, 1993.


\bibitem[Harding \& Lai(2006)]{Harding:2006qn}
  Harding, A. K., \& Lai, D. 2006,
  Rept.\ Prog.\ Phys., 69, 2631.

\bibitem[Harding \& Muslimov(1998)]{Harding:1998ma}
  Harding, A. K., \& Muslimov, A.~G. 1998,
   Astrophys.\ J., 508, 328.


\bibitem[Zhang \& Harding(2000)]{Zhang:1999ua}
  Zhang, B., \& Harding, A. K. 2000,
  Astrophys.\ J., 532, 1150.


\bibitem[Zhang et al.(2003)]{meszaros} Zhang, B., Dai, Z.G., Meszaros, P., Waxman, E., \& Harding, A. K. 2003, Astrophys. J., 595, 346.

\bibitem[Goldreich \& Julian(1969)]{Goldreich:1969sb}
  Goldreich, P., \& Julian, W.~H. 1969,
  Astrophys.\ J., 157, 869.


\end{thebibliography}

\end{document}